\documentclass[12pt]{iopart}

\newcommand{\ket}[1]{\ensuremath{\left| #1 \right\rangle}}
\usepackage[dvips]{graphicx}
\usepackage{verbatim}

\begin{document}

\title[Entanglement-based QKD at telecom wavelengths]{Fully automated entanglement-based quantum cryptography system for telecom fiber networks}

\author{Alexander Treiber$^1$,  Andreas Poppe$^2$, Michael Hentschel$^3$, Daniele Ferrini$^1$, Thomas Lor\"{u}nser$^2$, Edwin Querasser$^2$, Thomas Matyus$^2$, Hannes H\"{u}bel$^1$ and Anton Zeilinger$^{1,3}$}

\address{$^1$Quantum Optics, Quantum Nanophysics and Quantum Information, Faculty of Physics, University of Vienna,
    Boltzmanngasse 5, 1090 Vienna, Austria}

\address{$^2$Austrian Research Centers~GmbH~-~ARC,
  Donau-City-Str.~1, 1220 Vienna, Austria}

\address{$^3$Institute for Quantum Optics and Quantum Information,
Austrian Academy of Sciences, Boltzmanngasse 3, 1090 Vienna,
Austria}

\ead{anton.zeilinger@univie.ac.at}
\begin{abstract}
We present a quantum key distribution (QKD) system based on polarisation entanglement for use
in telecom fibers. A QKD exchange up to 50km was demonstrated in the
laboratory with a secure key rate of 550 bit/s. The system is
compact, portable
with a fully automated start-up and stabilisation modules for polarisation,
synchronisation and photon coupling allow a hands-off operation. Stable and reliable key exchange in a deployed optical fiber
of 16km length was demonstrated. In this fiber network we achieved over two weeks an automatic key generation
 with an average key rate of 2000 bit/s without manual
intervention. During this period, the system had an average entanglement visibility of 93\%, highlighting the technical level and stability achieved for entanglement-based quantum cryptography.

\end{abstract}


\maketitle


\section{Entanglement-based quantum key distribution}

Quantum cryptography, or more specifically Quantum Key Distribution
(QKD), is the most advanced quantum information protocol, concerning its prospect of commercialisation.
Since its first introduction 25 years ago with the BB84 protocol \cite{BB84} which, based on single-particle encoding ideas of Wiesner, did not use entanglement yet, the field has advanced enormously  both in theoretical developments and experimental realisations. Most significant was the proposal \cite{Ekert91} of entanglement-based quantum cryptography. There exists now
a whole variety of different physical implementations, each with their
own advantages and disadvantages. The interested reader is referred
to the overviews \cite{scarani,gisinQKD}.

When the QKD protocols using entanglement have been developed \cite{Ekert91,BBM92}, the use of this resource for
practical QKD devices was regarded as limited  because of its higher technological challenges. The onset of highly efficient down-conversion sources for entangled photon pairs
\cite{kwiat} has changed this situation significantly. Whereas the rate and distance of early
demonstrations of entanglement-based QKD \cite{jenn,qkdkwiat,gisinent} were still limited by the efficiency of the sources,
technological advances in the meantime \cite{shap,fed} have shifted the limitations to the single photon detectors.

Although difficult to realise, there are however reasons for preferring entanglement. This has been
recently highlighted by the development of device independent
security proofs \cite{devindependent}, which do not require an
\textit{a priori} trust of the QKD device.
We would also like to stress that developments for reliable and long
distance entanglement based QKD do not only benefit the
cryptographic community. Since entanglement is the resource of many
other quantum communication and quantum computation protocols, these would also profit from the
developments in entanglement-based QKD. We therefore are convinced that the technologies
developed for entanglement-based systems will benefit the quantum
information community as a whole.

To date, most entanglement systems employ free space links where the
wavelength of around 800nm is used. Inner-city links
were realized \cite{hughes,kurts,weihs} as well as long distance QKD
over 144km between two islands \cite{teneriffa}.
Early fiber-based implementations have also used a wavelength at 810nm \cite{jenn,bank},
but thereby heavily restricting the distances due
to the high absorption of optical fibers in this region.
The use of entanglement for QKD experiments performed at telecom
wavelengths has been very sparse \cite{gisinent30,gisinent50,yamQKD}, which
comes as a surprise given the multitude of entanglement distribution
experiments in telecom fibers over the last years
\cite{takesue,kumar,ox100km,yamamoto,lim}.

We present here an entanglement-based QKD system designed to work at
1550nm for optimal distribution in optical fibers. Our QKD system is
realised as a compact portable device offering reliable and stable
key generation. The start-up and alignment process is completely
automated. Stabilisation routines guarantee a hands-off and long-term
operation. In this paper we show the performance of our device both in the
laboratory where we achieved a 50km transmission and in the real
world where we demonstrated for the first time a long-term and high
fidelity QKD based on entanglement using deployed telecom fibers
\cite{gisinent50,yamQKD}.

\section{Description of the entangled QKD system}
\begin{figure}[t]
\begin{center}
\includegraphics[width=\textwidth,trim = 0mm 0mm 0mm 0mm, clip]{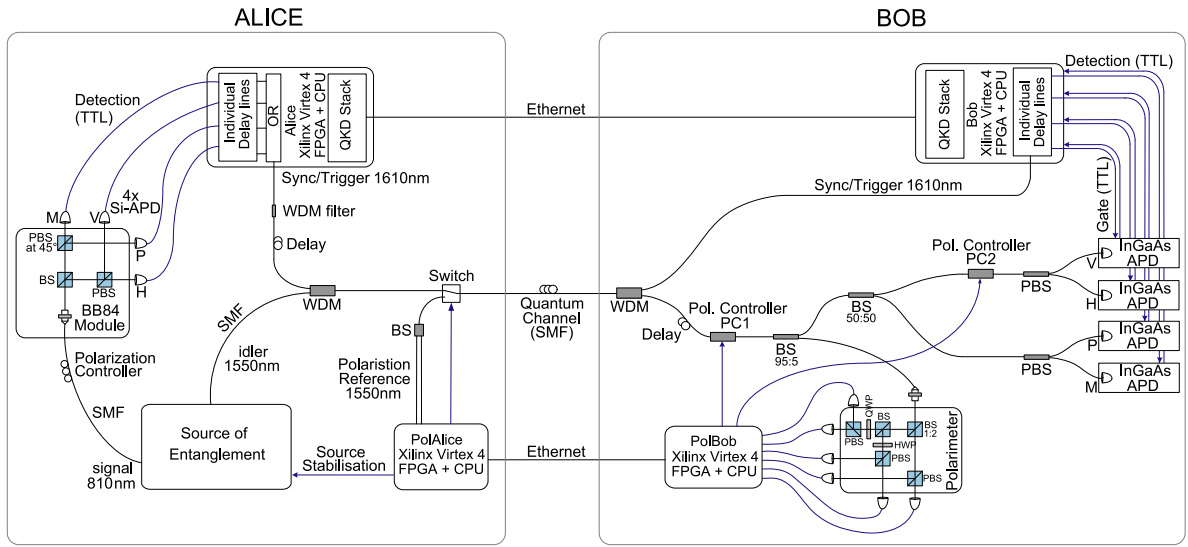}
\caption{(color online) Layout of the entanglement based QKD system. Alice contains the source of
polarisation entangled photons, a polarisation analysis module (BB84),
four Si-APDs and an electronic processing board. Furthermore,
components for generation of the trigger signals and part of the polarisation
control module are also included in Alice (PolAlice). On Bob's
side, an all-fiber BB84 module can be found together with four InGaAs
detectors. The receiver part of the
polarisation control (Polarimeter) and the necessary electronics are also located at Bob (PolBob). A fully automated polarisation alignment can be performed using the electronic polarisation controllers PC1 and PC2. The necessary connections between Alice and Bob for a secure key generation are a single mode fiber (SMF) and a classical network connection.}
\label{system}
\end{center}
\end{figure}

In this chapter an overview of the system, as depicted in figure
\ref{system}, is given together with an explanation of some of the
subsystems. The complete setup consists of the units ``Alice'' and ``Bob'', two
computers, a classical communication link and a quantum channel. A detailed description of the modules involved in
the automation and stabilisation is given in chapter \ref{controls}.

\subsection{Alice}

The core of Alice's unit is the polarisation-entangled photon source
emitting pairs with asymmetric wavelengths (810 and 1550nm)~\cite{ox100km}.
A 532nm cw-laser pumps two ppKTP crystals for type-I
spontaneous parametric down-conversion (SPDC). The two crystals are
poled for collinear emission of 810 and 1550nm. Arranged in the scheme of Kwiat \textit{et al.} \cite{kwiatcrystal}, the polarisation state after the two crystals is given by:

\begin{equation}
\ket{\phi}=\frac{1}{\sqrt{2}}\left(\ket{\rm{H_{810} \hspace{1mm}
H_{1550}}}+e^{i\phi}\ket{\rm{V_{810} \hspace{1mm} V_{1550}}}\right)
\end{equation}

The pair is separated inside the source using a dichroic mirror and each  photon
is coupled into a single mode optical fiber. At the 810nm side an
additional bandpass filter ($810\pm0.5$nm) is placed before the
fiber to limit the bandwidth of the photons.

The 810nm photons pass an in-fiber polarisation controller before
entering the BB84 module. Here the photons are recollimated onto a
balanced beamsplitter (BS) and then directed to two cube polarising
beamsplitters (PBS). The PBSs are rotated by $45^{\circ}$ relative
to each other to perform measurements in
the $0^{\circ}$/$90^{\circ}$ (H/V) and in the
$+45^{\circ}$/$-45^{\circ}$ (P/M) bases. The four output ports of
the PBSs are coupled into multimode fibers connected to an array of
four Si-APDs (SPCM-AQ4C from PerkinElmer). The TTL outputs from the detectors are fed into an electronics
board (Virtex-4 FX20 from Xilinx) for further processing. All events
are logged on the embedded CPU (PowerPC 405), where the raw
key is passed to a large buffer for subsequent key distillation.

Within Alice's unit, the optical fiber carrying the 1550nm photons is first
combined in a wavelength division multiplexer (WDM) with a trigger
line and the fiber then passes an optical switch necessary for the operation
of the polarisation control. The fiber is connected at the
front of Alice's case to the quantum channel, which consists of a single
mode telecom fiber, allowing the 1550nm photons to travel to Bob.

\subsection{Bob}

At Bob's unit, only telecom wavelengths are present, so almost all optical
components are fiber based. First, a WDM-demultiplexer separates the trigger signal from the single photons. After a delay line of 32m, the
photons pass through the electronic polarisation controller PC1 (PolaRite II from
General Photonics) after which 5\% are directed to the classical polarimeter.  The
main fraction of 95\% enters Bob's fiber based BB84 module, where a balanced BS (50:50)
randomly selects the measurement basis. One arm of the BS
leads directly to a PBS, whereas the other arm is attached to another
polarisation controller PC2 before a second PBS. This polarisation
controller allows to rotate the measurement axis of the second basis
relative to the first. The outputs of both PBSs are
connected to four InGaAs-APDs (id-201 from IdQuantique). The generated detector
signals are analysed by Bob's Virtex-4 electronics.

\subsection{Synchronisation}
\label{sync} Since the InGaAs detectors require a gate voltage when
a photon is expected, a synchronised timing channel is
needed. We opted for a time and wavelength
multiplexed optical trigger signal  which also runs over
the quantum channel to implement a one-fiber solution. Since the
SPDC is pumped with a cw-laser, timing information can only come from
the pair itself. Every time a 810nm photon is detected at Alice, a
strong optical pulse at 1610nm is generated. The sideband emission around 1550nm from the
laser diode is sufficiently suppressed by additional filters.
A WDM-multiplexer combines each 1550nm photon with its trigger pulse, which follows the single photon by
a few nanoseconds to reduce Raman scattering effects. Arriving at Bob, the trigger pulses are converted into  electronic signals which gate all
four InGaAs detectors. Since the conversion takes some time, the
single 1550nm photons are delayed in order to arrive at
the appropriate time at the detectors.

\subsection{19-inch packaging}

\begin{figure}[t]
\begin{center}
\includegraphics[width=\textwidth,trim = 0mm 0mm 0mm 0mm, clip]{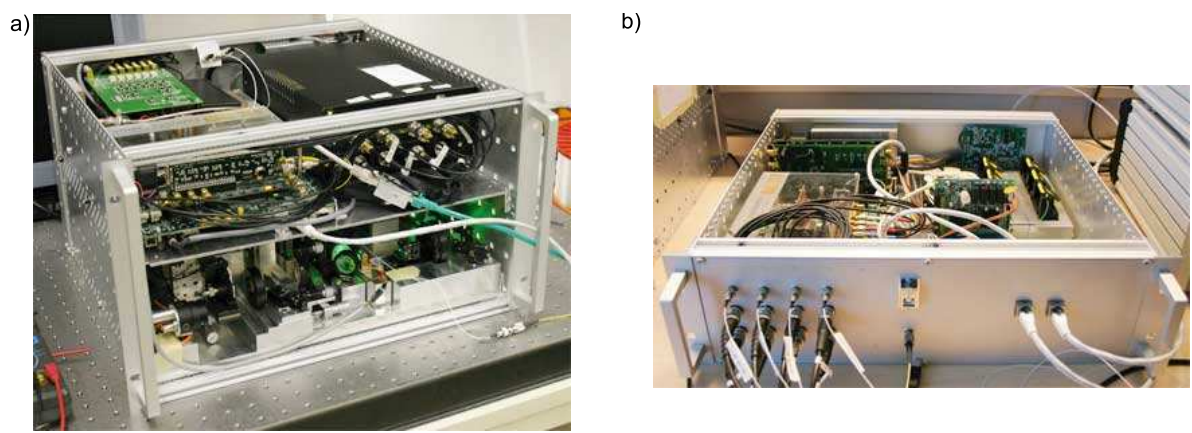}
\caption{(color online) (a) The ``Alice'' unit of the entangled
QKD link. The entanglement source is located at the bottom together with the BB84 polarisation analysis module. The Si-detector array and electronics for key generation, the trigger diode and reference diodes for the polarisation control are found on top. Connections are
provided by two ethernet ports and a coupler for the quantum
channel. (b) The receiver unit ``Bob'', containing a fiber based BB84 module, the electronics for key generation and a polarimeter for the polarisation control. Apart from the
coupler to the quantum channel and two ethernet ports, Bob has also
4 fiber outputs and 8 coaxial connectors. Those connectors are for
the four InGaAs detectors, which are placed and operated outside the
unit. Each unit is housed in an aluminium case (43x42cm) with
Alice standing at 6 Height Units ($\sim$ 26 cm) and Bob at 3
HU ($\sim$ 13 cm).} \label{Alicebob}
\end{center}
\end{figure}

The system was designed for use in installed optical networks, hence further
steps had to be taken to facilitate handling and installation of the
devices. It was decided to comply with a standard size and the units at
Alice and Bob
were required to fit into a 19-inch rack, as shown in
figure~\ref{Alicebob}.

The unit at Alice has
a two level structure, with the entanglement source and the BB84
module on the bottom to keep thermal effects low.
To enhance stability of the source, all optical elements were mounted
on a monolithic base plate ($42\times26$cm). On the top
level, the Si-APDs, the electronics for registering the counts and the polarisation control are
housed.

On Bob's side, the miniaturisation was helped by the fact that the
whole module is fiber based, with exception of the polarimeter. All
fibers are encased in  plastic  to prevent physical movement.
Due to the fact that the four InGaAs detectors are separate devices
and therefore rather bulky, an integration of the detectors inside
Bob was not possible. Four fiber ports on Bob's front
allow to connect each InGaAs detector to an output of the PBSs.
The corresponding trigger and detection signals are fed
through eight coaxial connectors.

\subsection{QKD protocol implementation and security considerations}
We implemented the BBM92 protocol
\cite{BBM92} for entanglement based QKD, a generalisation
of the original BB84 protocol \cite{BB84} applied to entangled states \cite{Ekert91}. Intuitively speaking, the key is just generated at both locations simultaneously. This elegantly circumvents the well known key-transfer problem of conventional cryptography.
In entanglement-based protocols, the perfect correlations in a shared entangled state are used to generate a secure key, as follows:
Both Alice and Bob keep track
of all their detection events together with a record of their respective polarisation
state measurement settings; Bob communicates to Alice which of the synchronisation triggers
resulted in a detection event on his side and also communicates his measurement basis; Alice deletes all entries
in which Bob did not detect a photon or where the bases did not match. Both parties share then a
sifted key, which is further processed
through the QKD protocol stack to yield a secure key. The software implementing
the QKD stack is located on two computers, one connected
to Alice and the other to Bob.
All classical communication between
them is routed over a standard network connection. Alternatively, the QKD stack can run directly on the embedded CPU for a single-chip quantum cryptography solution \cite{QCC}. This latter approach is currently limited to key rates of 1 kbit/s due to computational limitations of the embedded CPU.

In our implementation of the QKD stack, error correction and privacy
amplification are performed using the
 CASCADE algorithm \cite{cascade} and universal$_2$ class hash
functions \cite{PAU} respectively. The hash function is implemented using a T\"{o}plitz matrix
approach. We are secure against individual
attacks \cite{lutkenhaus} which have proven sharp
bounds \cite{indvattack}. Against coherent attacks, the proof by
Koashi and Preskill \cite{preskil} should be used to secure the
entanglement based QKD \cite{malo}, which would reduce the secret key by  $\sim8$\% for our typical QBER (3-4\%).
Although these proofs are strictly true only in the infinite key
limit \cite{keylimit}, we currently use sifted key blocks with a size of 32 kbit due to computational limitations.
To exclude a man-in-the-middle
attack, all communication between Alice and Bob is authenticated by
applying secure message authentication in the form of an evaluation
hash \cite{auth}.

In an actual realisation of a QKD device, side-channel attacks have
to be addressed (see section \ref{dsync}). These attacks exploit the non-ideal parts of a QKD
system and have various origins \cite{kurtsieferattack,
maloattack,markarov}. In our case the detection efficiencies of the
single photon detectors are not perfectly matched. A better
selection of detectors and bias adjustments can however be
implemented.

Due to the nature of spontaneous down-conversion, multi-pair events
are generated, which might leak information to a potential
eavesdropper \cite{multi}. To counter this attack a random
measurement result should be assigned whenever double detection
events occur~\cite{detectorsqash}. Our electronics only process the
first detection event coming from the detector array, before
blanking all detectors with an artificial dead time (300ns). This ensures a
quasi randomisation of double events, since the timing jitter of the
detectors ($\sim$500ps) is much larger then the coherence time of
the photons~($\sim$2ps).

\section{Results from the laboratory}

With the QKD system ready in its compact form, we measured the
performance within the laboratory environment. Firstly, we
investigated the qubit error rate (QBER) and the secure key rate as function of the laser
intensity to determine the suitable pump power. Secondly, we tested
the long distance capability of the system with fiber spools.

\subsection{Secure key rate and qubit error rate}

\begin{figure}[t]
\begin{center}
\includegraphics[width=\textwidth,trim = 0mm 0mm 0mm 0mm, clip]{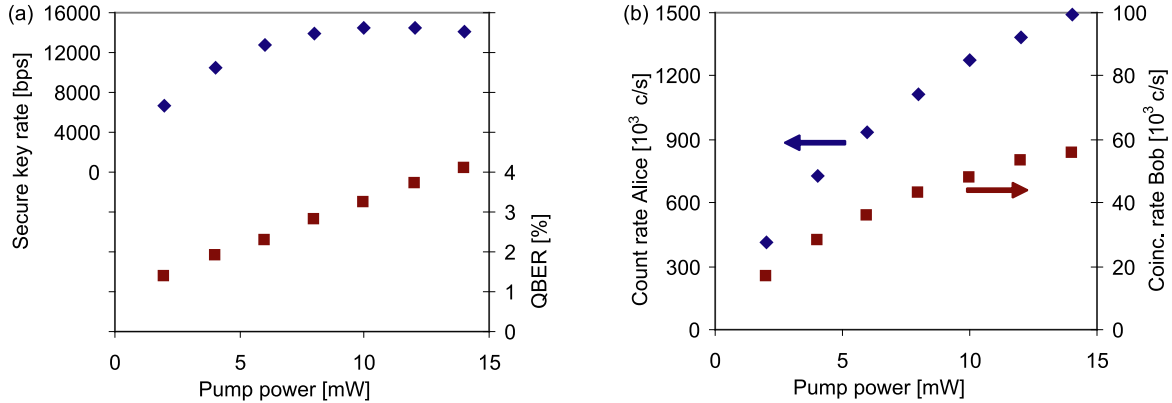}
\caption{(color online) (a) The secure key rate (blue diamonds) and the QBER (red squares) produced by the entanglement based QKD device.  The data was taken at short distance and
varying pump power. (b) The total count rate of the four Si-APDs at Alice (blue diamonds) and the total
coincidence rate  detected by the four InGaAs-APDs at Bob (red squares) recorded at varying pump power.} \label{power_qkd}
\end{center}
\end{figure}

The first test of the QKD system was performed at various pump power settings.  The
measurements presented in this section were taken with
Alice and Bob connected only by a few meters of optical fiber.
Data was collected at power settings ranging from 2 to 14mW and averaged
over 10 minutes. At 2mW, Alice produced a trigger rate of $415000$
counts/s. Bob's total coincidence rate for this setting was $16600$
c/s, yielding a coincidence probability of $\sim4 \%$. The overall
losses of coincidences can be attributed to the following:
conditional fiber coupling of the 1550nm photons at the source ($50\%$),
transmission at Alice ($75\%$), transmission at Bob
($65\%$) and a
detector efficiency of ($15\%$). The QBER measured at this power
setting was 1.4\%, yielding a secure key rate after sifting, error
correction and privacy amplification of 6500 bit/s. Figure
\ref{power_qkd}(a) summarises the values for the other power settings.

One can directly see in figure \ref{power_qkd}(a) that the secure
key rate first rises with the pump power, then has a maximum around
10mW and finally decreases for even higher power. The reason for
this behaviour is twofold. Firstly, the raw count rates are not linear
with the intensity but show a saturation behaviour, see figure
\ref{power_qkd}(b). This is readily understood from detector dead
times and also from the gated operation of the InGaAs detectors. The
minimum time difference between two consecutive gates is 250ns. Triggers which arrive
 during this time are ignored. Since we did not
want to overload our system, the dead time of Alice's detectors was
electronically enlarged to 300ns, suppressing the unusable events
already at Alice. This leads to the saturation of Alice's counts and
to the sub-linear increase in Bob's coincidences with intensity. Note that the
coincidence probability remains the same over the whole power range.
The second and more important reason why the key rate is actually
falling at higher power is the rise in QBER. The QBER increases
linearly with the pump power, as shown in figure \ref{power_qkd}(a), due to multi-pair emissions from the
down-conversion process.
Multi-pairs from cw-pumping are largely unrelated and give therefore
uncorrelated results \cite{multi}. Since the multi-pairs grow quadratically with pump power as
opposed to the linear increase of the single pair coincidences, the
QBER increases linearly, too.

In order to verify this assumption, we
decomposed the QBER value into three parts: system error, dark count
error, and accidental coincidence error. The system error comprises
all the non-ideal components of the system, such as polarisation
splitters and polarisation alignment, but also the imperfection of
entanglement from the source. The dark count error is solely caused by coincidences between a dark count on Bob's InGaAs-APDs and a trigger pulse (detection event at Alice). The contribution of Alice's Si-APD dark counts to the trigger is however negligible. The accidental coincidence error comes
from multi-pair generation within the
detection window ($\sim1.5$ns).

\begin{figure}[t]
\begin{center}
\includegraphics[width=11cm,trim = 0mm 0mm 0mm 0mm, clip]{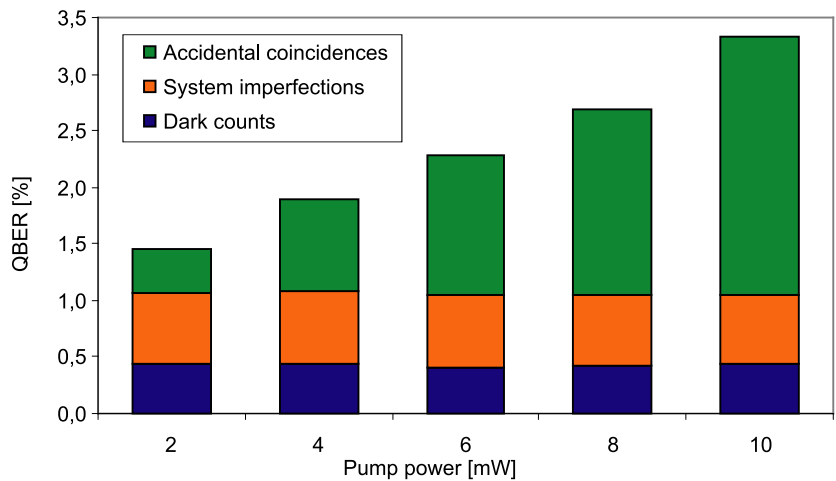}
\caption{(color online) The measured error (QBER) of the entangled QKD system recorded at different power settings. The QBER is analysed coming from three error sources:
The dark count and system QBER, caused by the dark counts of the InGaAs detectors and the non-ideal components of the system respectively and the accidental coincidence
QBER arising from multi-pair production in the down-conversion process. Whereas the former two error contributions remain stable, the accidental coincidence
QBER rises linearly with the pump power.} \label{qberpower}
\end{center}
\end{figure}

Figure \ref{qberpower} shows each error component for the power
range of 2 to 10mW, with the linear increase of the overall QBER
clearly visible. Furthermore, the rise can be fully attributed to the
increase in the accidental count rate. Both the system error and the
dark count error remain constant for the investigated power levels. The intrinsic
error of the system is very low indeed, contributing only 0.6\% to
the total QBER. Detector dark counts already play a role at short distance,
adding 0.4\% to the QBER. The error due to the accidental counts
increases from just below 0.5\% to more than 2\%, limiting the use of
larger pump powers. To decrease the accidental rate, low jitter
detectors and electronics need to be used together with narrower
coincidence windows, both beyond the scope of this work.

To conclude this analysis, we recognise that the optimal power level will depend on the actual fiber length,
but decided to leave the intensity at 6mW for all subsequent measurements.
This setting is a compromise between a low starting QBER and the anticipated
rise of the error when longer fibers are included, due to the prominent role of dark counts.

\subsection{Long distance measurements}

\begin{figure}[t]
\begin{center}
\includegraphics[width=12cm,trim = 0mm 0mm 0mm 0mm, clip]{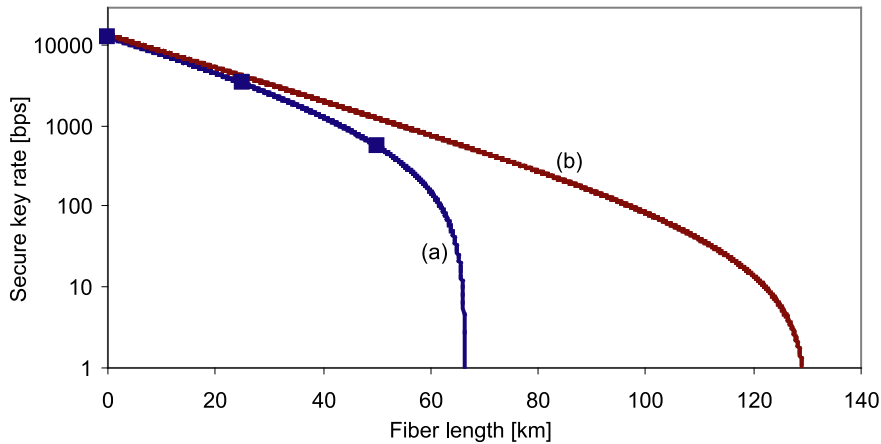}
\caption{(color online) Measured secure key rate (blue squares) of the entangled QKD system as a function of fiber length at 0, 25 and 50km. The curves are predictions for the secure key rate with distance obtained from eq. (\ref{key}). (a) shows the prediction for our actual system with a measured dark count rate per InGaAs detectors of $d=80$~c/s. (b) shows a best effort prediction using the dark count rate of the best InGaAs detector ($d=4$ c/s).
The following experimental parameters were used
for the model: $\alpha = 0.204$dB/km, $c_0=9000$ c/s, $a_0=240$ c/s
and $f(e) \approx $ 0.17.}
\label{longdistance}
\end{center}
\end{figure}

The long distance capability of the QKD system was investigated in
the laboratory using spools of optical fiber. Since the
bandwidth of the 1550nm photons is around 3nm, chromatic dispersion
of a standard fiber would limit the distance. We therefore used non-zero
dispersion shifted fibers with a chromatic dispersion of 4 ps/km/nm
(TrueWave RS fiber from OFS). QKD measurements were performed at distances of 0,
25 and 50km with a duration of 60 minutes each.  With 6mW of pump
power, the values for the key rate and QBER at 0km are 12500 bit/s
and 2.3\% respectively, as shown in figure \ref{longdistance}. At
25km, the QBER increased to 3.3\% and the obtained key rate was 3300
bit/s. At the longest distance of 50km, a secure key rate
of 550 bit/s was still observed with an overall QBER of 6\%. The strong increase in
QBER with distance confirmed our previous decision to start with a
low QBER in order to accommodate possible rises. At larger distances the
dominant contribution to the QBER is due to detector
dark counts and no secure key could be produced at distances much
beyond 50 km. To show that our rates are dark count limited we
devised a simple model for our experimental data.

The average coincidences and accidental coincidences decrease
exponentially with the length of the fibre ($l$):
\begin{equation}
c(l) = c_0\times10^{-\alpha \cdot l / 10} \qquad a(l) = a_0\times10^{-\alpha
\cdot l / 10}
\end{equation}
where $c_0$ and  $a_0$ are the coincidence and accidental
coincidence rate (from multi-pairs only) at 0km per detector respectively, and $\alpha$ is the
attenuation per kilometer.
The total coincidence rate is therefore $4\,c(l)$ and the total background rate is $4\,a(l)+4\,d$, where $d$ is the average dark count rate per InGaAs detector.

The QBER is defined as \cite{gisinQKD}:
\begin{equation}
QBER = \frac {n_{false}}{n_{true}+n_{false}}
\end{equation}
where $n_{true}$ and $n_{false}$ are the true and false bits of the key respectively.
After Sifting, the true bits are given by half of the coincidence rate
($2\,c(l)$) and a quarter of the total background rate ($a(l)+d$), whereas the false bits are given by just a quarter of the total background rate ($a(l)+d$).
Hence, the error from the accidental coincidences and dark counts reads as:

\begin{equation}
e_{noise}(l) = \frac{a(l)+d}{2\,c(l)+2\,a(l)+2\,d}
\end{equation}

The final secure key rate after sifting, error correction and
privacy amplification is given by:
\begin{equation}
\label{key}
k_{sec}(l) = \frac{4\,c(l)}{2}(1-\tau(e)-f(e) \cdot h(e))
\end{equation}

where the factor $1/2$ accounts for sifting, $e$ is the overall QBER
($e=e_{noise}(l) + 0.6\%$ for the system QBER), $h(e)$ is the
binary entropy function, $f(e)$ is the overhead of CASCADE with respect
to the Shannon limit and $\tau(e)$ is the fraction discarded by privacy
amplification \cite{lutkenhaus}. In our case the overhead of CASCADE ($f(e)$) was measured to be 1.16 at 3\% QBER and 1.18 at 5\% QBER.

The average dark count rate per InGaAs detector was measured to be
$d=80$~c/s and the expected key rate from equation (\ref{key}) is
plotted in figure \ref{longdistance}(a). The prediction of the model
is in very good agreement with the experimental data points and
hence confirm that our long distance measurements are indeed dark
count limited, since the dark count rate (80 c/s) exceeds the
accidental coincidence rate (74 c/s)  for distances beyond 25km. The
question arises what distances could be achieved with better
detector technology. Our best InGaAs detector has a measured dark
count rate of only 4 c/s for the given trigger rate of 950000 c/s
and coincidence window of 1.5ns. The prediction for the key rate in
a scenario with four low-noise detectors ($d=4$ c/s) is given in
figure \ref{longdistance}(b). A distance of over 100km would
therefore be possible with key rates of tens of bit/s at 100km.

Polarisation Mode Dispersion (PMD) was thought to prevent such
distances for polarisation encoding, however modern telecom fibers
are optimized to typical PMD link design values of smaller than
$0.07 \mathrm{ps/}\sqrt{\mathrm{km}}$. Only after 1200 km would the
PMD overcome the coherence time of our entangled photons (2.5ps) and
lead to decoherence. Previous investigations \cite{ox100km} revealed
a 100km fiber link to have no adverse effects on polarisation
entanglement. Narrowband sources of entanglement \cite{kistadwdm}
can be used to reliably transmit polarisation qubits with coherence
times (10ps) much larger than the PMD value of a 1000km link (2ps).
We conclude that even with current InGaAs technology an entanglement
based QKD link of 100km in optical fibers can be realised.

\section{Automation and stabilisation modules}
\label{controls} As was shown in the previous chapter, long distance
entanglement based QKD with high key rates is possible. For a practical QKD system, one requires in addition that human intervention should be minimised.
Ideally, the system should be completely automated and achieve a stable and reliable long-term key exchange.
For the system presented here it is therefore necessary to distribute photonic entanglement with a stable
rate and high visibility. In particular, this requires:

\begin{itemize}
\item Stable production and coupling of photons into the quantum
channel
\item Stable synchronisation to identify photons from the
same entangled pair
\item Automated alignment of the entangled state
\item Compensation of polarisation drifts in the quantum channel

\end{itemize}

Most of these requirements are not necessarily specific to our QKD
prototype but apply to all quantum information protocols relying on
entanglement distribution, some even to other protocols as well.
In the following subsections, we describe the implemented modules meeting the requirements mentioned above.

\subsection{Source stabilisation}
\label{stab}

The entanglement source is based on free-space optics and hence prone to mechanical drifts due to temperature fluctuations.
Even small drifts will lead to misalignment and
loss of fiber-coupled photons. Therefore an active stabilisation
procedure was implemented. The fiber couplers (fiber tip with a fiber
collimation lens) for the photons have been fixed on
piezo-motor driven optical mounts (Agilis AM-M100 series from
Newport). These mounts can tilt the fiber coupler along two axes in order to
stay aligned. Another motorised mirror mount was installed just
after the laser to compensate beam wander.  All six piezo
axes are driven by the electronics at Alice and an algorithm
maximises the count rates. The source stabilisation is not a fast procedure, as due to the long
averaging times involved (10 seconds per point), it takes several
minutes to complete a full cycle (all 6 channels). This does however
not decrease the key rate, since the QKD continues while the source stabilisation is active.

\subsection{Delay synchronisation}
\label{dsync}
As outlined in section \ref{sync}, the  synchronisation channel is essential
to trigger the InGaAs detectors at Bob. To have more flexibility and
control over the exact
synchronisation, electronic delay lines  were added at Alice and Bob (see figure \ref{system}).
The delay lines can be addressed individually and set in steps of 10 ps for a
maximum delay of 10ns.

On Alice's side the delay lines are found between the outputs of the Si-APD detectors and the input of the FPGA electronics. This arrangement allowed us to compensate detector-dependent response-times and to
  match the optical trigger signal to the single photon arrival irrespective of which of the four
Si-APD detectors fired and therefore preventing a potential side channel attack
\cite{kurtsieferattack}.

The delay lines at Bob were inserted between the generation of the electronic trigger signal
and the gate inputs on the four InGaAs detectors. With
those delay lines we could precisely synchronise the gate of
each InGaAs detector to the single photon, therefore also preventing a
time-shift attack \cite{maloattack}.

An automated synchronisation routine is implemented for Bob's delay lines. At the start, Bob sets all delays to minimum and then increases the delays stepwise to the maximum value. The coincidence rates are monitored during the scanning process and the optimal delay for each detector is obtained. Since temperature fluctuations also affect the delay circuits, the routine has to be repeated periodically.

The width of the coincidence peak is given by detector
jitter and in longer fibers by the chromatic dispersion. The
smallest gate width on the id-201 detectors leads still to the detection of
more dark counts and accidental coincidences than necessary. An
additional coincidence-window can be defined electronically to fit the
arrival-time spread of the photons tightly.

\subsection{State alignment}
\label{statealign}

The purpose of this routine is to perform an automatic polarisation
alignment on the desired maximally-entangled state $\ket{\Psi^{-}}$ to minimise the QBER.
Whereas on Alice's side a free-space arrangement of the BB84 module guarantees
a permanent and fixed $45^{\circ}$ rotation between
the two measurement bases H/V and P/M, on Bob's side
the BB84 module is fiber based and  no absolute reference exists
between the two arms. Since temperature changes will influence both arms independently
and any initial reference will be lost, we implemented an automated state alignment procedure using the two polarisation controllers at Bob (PC1 and PC2 in figure \ref{system}).
Independent bases control is achieved with the first controller acting on both bases and the second controller acting only on the H/V basis. At the start of
the procedure, Alice only triggers with events coming from her P-detector (see figure \ref{system}). Bob then uses the first polarisation controller (PC1) to reduce
the coincidence rate in his P-detector to a minimum
($\ket{\Psi^{-}}$).
 After the P/M basis is aligned, Bob
uses the second controller (PC2) to align the H/V basis by minimising the
coincidences at his V-detector, while Alice triggers with her V-detector only. The whole alignment process takes about 3 - 5 minutes for
each basis (10 seconds averaging time for each measurement).

\subsection{Polarisation control}

Using standard optical fibers as a quantum channel for polarisation encoded photons
 has the disadvantage of arbitrary unitary polarisation transformation.
The birefringence of the
fiber causes arbitrary rotations depending on environmental factors
like temperature and mechanical stress. These random and dynamic
changes in birefringence, which would lead to an unacceptable high
QBER, need to be actively compensated for.
The state alignment procedure, detailed in section \ref{statealign},
could in principle correct the drifts in the quantum channel but it is slow and would interrupt the QKD for too long.
We therefore devised a fast and active control module compensating the dynamic polarisation rotation of the quantum channel.

Our implementation of the polarisation control relies on time
multiplexed reference pulses at the same wavelength as the
entangled photon. A wavelength multiplexed version has also been
shown recently \cite{WDMpolcontrol}. When in operation, Alice sends strong pulses
consecutively polarised in H and P through the fiber to
Bob, where their polarisation is analysed. If a difference from a
preset polarisation state is found, an algorithm controlling the
first polarisation controller (PC1) will minimise the deviation.
Two reference pulses at non-orthogonal polarisations are needed, as simply fixing a single
polarisation on the Poincar\'{e} sphere leaves the phase still undetermined. A second non-orthogonal polarisation state defines the phase, and hence all polarisation states on the sphere.

The hardware needed to implement the polarisation control is depicted in figure \ref{system}. On Alice's side,
the electronics control two laser diodes, whose outputs are combined in a BS maintaining a relative polarisation rotation of $45^\circ$ between them.
Alice (FPGA ``PolAlice'') also controls the optical switch (Free-X from CIVCOM) which
connects the quantum channel to either the entanglement source or to the
diodes of the polarisation control. The switch also interrupts
the trigger signals from Alice when the polarisation control is
active, preventing a saturation of Bob's detectors. At Bob, a 95:5
BS diverts a fraction of the signal to a classical
polarimeter analysing the light in the two linear H/V and P/M and
in the circular R/L bases. Photo diodes at each output port measure
the components along each basis and the incoming polarisation state can be
reconstructed from the evaluated Stoke's vectors \cite{polbook}.

\begin{figure}[t]
\begin{center}
\includegraphics[width=\textwidth,trim = 0mm 0mm 0mm 0mm, clip]{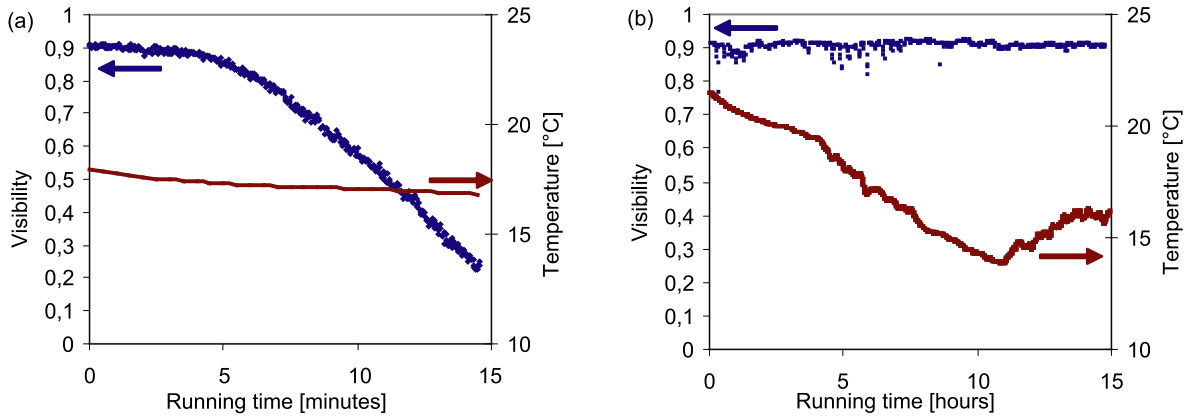}
\caption{(color online) (a) Evolution of the average visibility (blue diamonds) in the H/V and P/M basis for an entanglement distribution over 25km fiber
 without polarisation control. The temperature  at the fiber spool was measured at the same time (red line). (b) Evolution of the visibility (blue squares) over a whole night in a measurement with the active polarisation control. The highly varying temperature  at the fiber spool was again recorded (red line). Spikes in the visibility  are
caused by hysteresis effects of the polarisation controller~\cite{genphot}.} \label{pol}
\end{center}
\end{figure}

The polarisation control was tested in the laboratory using a 25km fiber
spool at varying temperatures, as shown in figure \ref{pol}. We used our source of polarisation
entanglement to measure the average polarisation visibility of the
entangled state in the H/V and P/M bases after transmission of the 1550nm photon through the fiber.
The visibility in one basis is defined as $V =
\frac{Max-Min}{Max+Min}$, where $Max$ is the maximal coincidence
rate as obtained for orthogonal polariser settings for the 810 and
1550nm photons and $Min$ is the coincidence rate at parallel
settings. Figure \ref{pol}(a) shows the measured
 visibility as a function of time (5 seconds averaging) together with the temperature of the spool. Without
polarisation control, even a small change in temperature
($\sim1.2^{\circ}$C) is enough to completely destroy the
polarisation correlations, which shows the
 high susceptibility of the fiber to external changes.
 The experiment was then repeated over a whole night (15 hours) with the polarisation control module.
 The constant visibility of the
 entangled state around 90\%, as shown in figure \ref{pol}(b), proves that the active polarisation control indeed compensates polarisation drifts even during large temperature changes ($7^{\circ}$C).

\subsection{Management module}

The management module software (MM) was developed to guarantee a
seamless coordination and operation of the individual control
routines. It manages the QKD system during the automated start-up and hands-off operation, and
reacts to different error scenarios.
During the normal QKD operation, the MM will call each of
the routines sequentially to keep the whole system stable and aligned. The
typical time for a single cycle, including the source stabilisation, time synchronisation
and the polarisation control, is about 10 minutes, after which the cycle
is repeated. Within a cycle, the QKD is only interrupted for about 5 seconds due to the polarisation control.

Interruptions of the normal cycle can be caused by several
error interrupts detected by the MM: if the optical fiber linking Alice and Bob is subject to violent
changes (temperature or mechanical stress), the QBER will quickly rise since
the polarisation control is not always active. The MM can however
identify this error and react by starting the
polarisation control. If the QBER rises slowly and is not reduced by the polarisation control, then the MM starts the state realignment procedure, since it is likely that parts of the fiber based BB84 module have drifted.

\begin{figure}[t]
\begin{center}
\includegraphics[height=8cm]{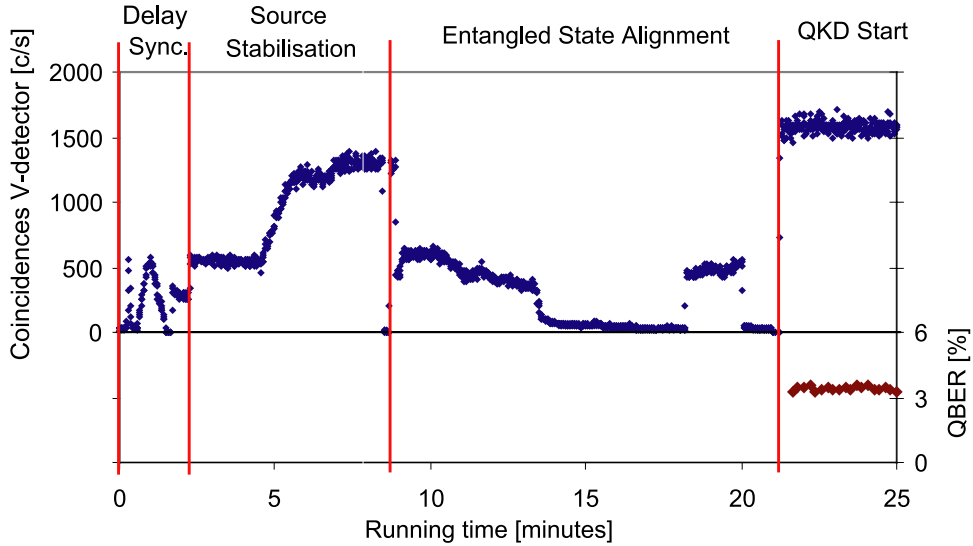}
\caption{(color online)  Automated start-up sequence of the QKD
system after being transported. The coincidence rate (blue diamonds) of one InGaAs
detector (V) is plotted over time to identify the different stages. At first, the system scans for the optimal synchronisation delay (identified by the spike), then the source stabilisation maximises the coincidence rate and finally the state is aligned in two subsequent cycles by minimising the V coincidences. After completion of the system start-up (21 min) the QKD is started. In this last section the QBER is also shown (red diamonds). }
\label{MMstart}
\end{center}
\end{figure}

An automated start-up sequence with a 25km fiber spool between Alice
and Bob is depicted in figure \ref{MMstart}. The coincidence rates on the V-detector at Bob are
shown as a function of time to identify each stage of the start-up. At
first, the system searches for the appropriate delay using the delay
synchronisation module (section \ref{dsync}). Then the source stabilisation module (section \ref{stab}) starts and in this
case increases the coincidence rate from 500 to about 1200 c/s within
5 minutes. After that, the state alignment module (section \ref{statealign}) minimises the
coincidences of the detector to align the H/V basis. The system is
now fully aligned and the QKD starts with a QBER of about 3.5\%.
The whole start-up took about 20 minutes after transportation by car and a
complete reinstallation of the system. The time for a routine start-up (after a temporary shutdown) takes only about 10 minutes.

\begin{figure}[t]
\begin{center}
\includegraphics[height=6.5cm,trim = 0mm 0mm 0mm 0mm, clip]{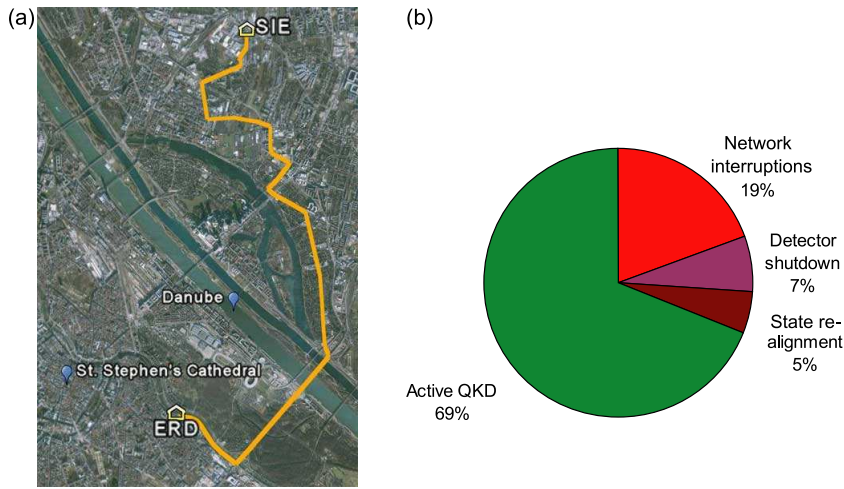}
\caption{(color online) (a) Satellite image
($^{{\tiny{\copyright}}}$Google) of Vienna showing the path of the
underground fiber (16km) between the locations SIE and ERD, over
which the entangled QKD link was operated. (b) Chart of the QKD
availability during the two-week network phase. The chart shows the
fraction of actual key generation (active QKD), together with the
interruptions caused by SECOQC maintenance, system realignments and
detector shutdowns due to high temperatures.} \label{map}
\end{center}
\end{figure}

\section{Long-term results from the SECOQC network}

\begin{figure}[t]
\begin{center}
\includegraphics[width=13cm,trim = 0mm 0mm 0mm 0mm, clip]{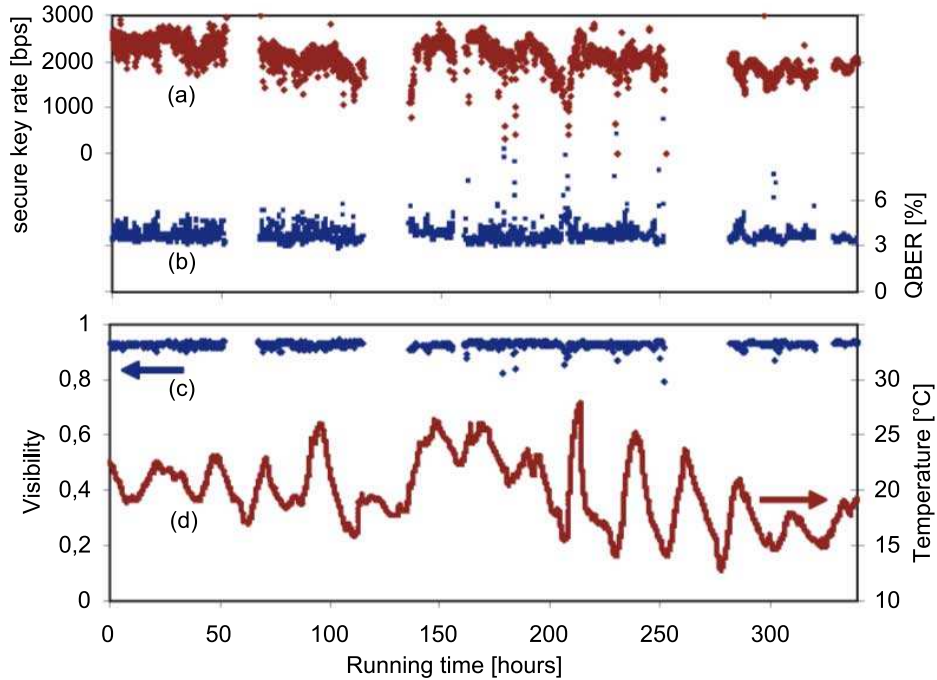}
\caption{(color online) Data obtained during the two-week network
phase (336 hours) using the 16km long deployed fiber. (a) Secure key
rate (red diamonds) and (b) QBER (blue squares) of the entanglement
based QKD link. Fluctuations in the key rate are caused by
temperature dependencies of the InGaAs detectors. (c) Average
polarisation visibility (blue diamonds) of the entangled state
measured during the network phase. (d)~The room temperature (red
line) at the ERD office (Bob) as measured throughout the network
phase. The large day/night changes are caused by an open window.}
\label{secoqc_long}
\end{center}
\end{figure}

The final test of our entanglement-based QKD device came with the
implementation of the link into the telecom fiber network provided by Siemens Austria as used by SECOQC \cite{seco}.
Alice's unit was placed in an office room at the Siemens headquarters (SIE)
in Vienna, Austria. Bob's unit was installed in another part of the city at
a branch office of Siemens (ERD). The two locations were connected
via a standard telecom fiber (G.652) of 16km length with a
total attenuation of 4.1dB. In figure \ref{map}(a), the actual path of
the fiber between SIE and ERD is depicted. Note that the fiber is
crossing the Danube river by means of a bridge, runs along major
train tracks and along a motorway. External influences on
the fiber were however not the only problem. Since the offices which housed
the QKD devices and several computers
 for the network had no air conditioning, a more archaic method had
 to be used to ``control'' the temperature. An open window allowed a
 capping of the temperature to prevent devices from shutting down, but
 resulted in large temperature variations inside the room. In figure
 \ref{secoqc_long}(d), the actual room temperature at ERD is shown for the whole phase
 of operation. The difference between day and night
 cycles are clearly visible with fluctuations in temperature up to $12^{\circ}$C.

The data in figure \ref{secoqc_long} shows both the secure bit rate
and the QBER of the link during the 336 hour run (two weeks). Each
point was averaged for about 1 minute. The key rate, shown as (a), lies around
2000 bit/s for the whole duration. Fluctuations in the key rate
are caused by unexpected temperature dependencies of Bob's detectors (e.g. at hour 220 in figure \ref{secoqc_long}). Note that the trigger rate at Alice (not shown here) was stable at 750000 c/s
throughout the two weeks. Bob's total coincidences decreased from
8500 c/s to 6800 c/s during this time, resulting in the decline of the key
rate. We believe that this is also linked to the
temperature dependence of Bob's detectors.

The QBER, shown as (b), remains very stable over the whole duration of the experiment
with an average value of 3.5\%. The apparent noise in the QBER plot is not caused by
random fluctuations but is due to the polarisation drifts not
compensated by the polarisation control procedure. As explained in
section \ref{statealign}, certain drifts can only be corrected using
the state alignment procedure. It is triggered when the QBER grows by an
additional percentage point from its initial value. Depending on the
temperature changes the state alignment module operates at intervals ranging from
30 minutes to several hours. In terms of entanglement visibility, shown as (c),
we achieved a two-week entanglement distribution with an average visibility of
93\%. For 99.9\% of the active time, the visibility was larger than
90\%, showing the high reliability of the entanglement distribution.

For two weeks, the QKD system was running continuously without
manual interference. The only times when no keys were produced was
during interruptions of the QKD network, where the link was embedded.
These periods lasted usually a day or so and
can be clearly seen in figure \ref{secoqc_long}. After the restart
of the network, the QKD link was also started automatically and both
QBER and key rate resumed from their previous values.  Out
of the total 336 hours of recorded data, the QKD was active for
232 hours (69\%), as depicted in figure \ref{map}(b). The
non-operating times were mainly caused by network interruptions,
which amounted to 64 hours (19\%) and detector
shutdowns accounting for 23 hours (7\%). The remaining 17 hours (5\%)
were spent realigning the state. It is reasonable to assume that with
a mature network infrastructure and air conditioned rooms, the
operational up-time could easily be 95\%. A temperature stabilised environment would
also reduce recurrence of the automated state alignment, which could bring the overall up-time to around 98\%.

\begin{figure}[t]
\begin{center}
\includegraphics[height=7cm,trim = 0mm 0mm 0mm 0mm, clip]{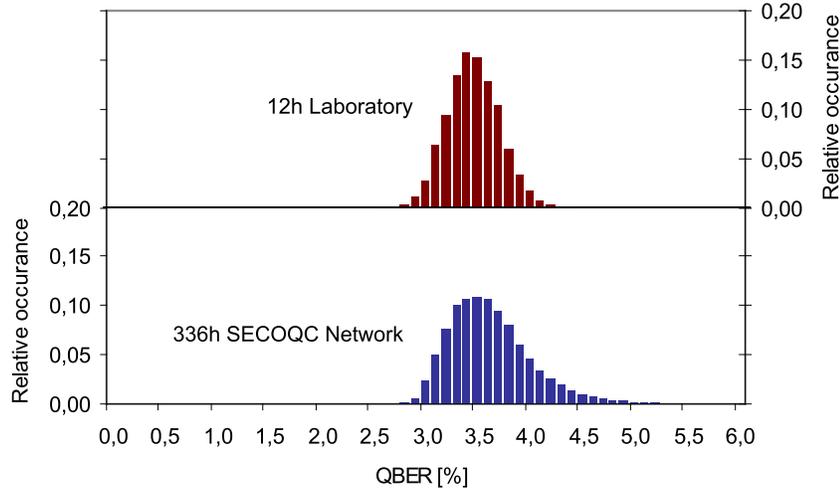}
\caption{(color online) Histogram of the measured QBER (a) during a typical
12h-run under laboratory conditions using a 25km fiber spool and (b) during the complete two week
network phase (336 hours) using the 16km long deployed fiber. } \label{qber}
\end{center}
\end{figure}

The high degree of robustness of the QKD system can be seen in
figure \ref{qber}.  The QBER histogram shown on top was recorded during the 12-hour
run in the laboratory using a 25km fiber, while the lower histogram
incorporates the QBER of the 336-hour run using the deployed fiber of the network. The two
distributions have nearly the same spread with 0.6\% FWHM in the
laboratory compared to 0.8\% FWHM in the network. The QBER
distribution in the network is slightly skewed to higher values,
since we allow for a QBER increase due to indoor temperature changes before
the state alignment corrects the drift. The similarity of the
distributions indicates however that our system can manage perfectly the
change from a stable laboratory environment to long-term operation
in a deployed fiber network.

\section{Conclusion}

We presented a QKD system based on
polarisation entanglement for use in telecom fibers. The whole system is compact and can be
used in standard 19-inch racks. We operated the system at a transmission
distance of up to 50km in optical fibers with a final secure key rate of 550 bit/sec.
Simulations show that with current InGaAs detector technologies
distances of up to 100km would be possible.

In addition, a variety of control and stabilisation modules were
combined for the first time to demonstrate a stable distribution of
entanglement with high purity and hence a reliable QKD implementation. With those extra modules, it was possible to realise a complete hands-off operation of the  whole QKD system. During two weeks and without any manual intervention, we measured
a key rate of 2000 bit/s over a 16km long deployed fiber with a QBER of 3.5\%.

These results also show that stable entanglement distribution
in optical fibers is possible over long periods of time. The high
visibility (93\%) of the entanglement and its reliability (in 99.9\% of the time higher than 90\%) makes application such as QKD feasible and hopefully stimulates the application of other
quantum communication protocols in optical fiber networks.

Since an entangled QKD system needs single photon detection, in contrast to other QKD implementations, at Alice \textit{and} Bob, the clock rate is currently restricted to the MHz region and thus limits the key rate.
However our experiment proves that entanglement-based quantum cryptography can be operated at least as reliably as simpler systems based on weak pulses. Thus the technological challenges which entanglement-based systems were facing initially have successfully been overcome. Such systems therefore offer a very secure and advanced approach to quantum key distribution.

\ack We would like to thank Bernhard Schrenk, Thorben Kelling,
Fotini Karinou and Bibiane Blauensteiner for their initial
contributions. We would also like to thank Michael Meyenburg and
Roland Lieger for the software support of the Virtex-4 FPGA boards as well as Sebastian Sauge
and Anders Karlsson for the loan of an InGaAs detector and discussions on entanglement distribution. We also acknowledge
Siemens Austria for the usage of their fiber network. This work was
supported by the European Commission through the integrated projects
SECOQC (IST-2003-506813) and QAP (015846) and by the Austrian Science
Foundation FWF (TRP-L135 and SFB-1520).

\end{document}